\documentstyle[prl,aps,epsfig,graphics,twocolumn]{revtex}

\begin{document}
\def\beq{\begin{equation}}
\def\eeq{\end{equation}}
\def\bea{\begin{eqnarray}}
\def\eea{\end{eqnarray}}
\def\oupb{UPB\ }
\def\pb{PB\ }
\def\eps{\epsilon}
\newcommand{\ket}[1]{| #1 \rangle}
\newcommand{\bra}[1]{\langle #1 |}
\newcommand{\braket}[2]{\langle #1 | #2 \rangle}
\newcommand{\proj}[1]{| #1\rangle\!\langle #1 |}
\newcommand{\eins}{\mbox{$1 \hspace{-1.0mm}  {\bf l}$}}
\newcommand{\ba}{\begin{array}}
\newcommand{\ea}{\end{array}}
\newtheorem{theo}{Theorem}
\newtheorem{kor}{Corollary}
\newtheorem{defi}{Definition}
\newtheorem{rema}{Remark}
\newtheorem{lem}{Lemma}
\newtheorem{exam}{Example}
\newtheorem{prop}{Property}

\draft

\title{Separability and entanglement in ${\cal C}^2\otimes{\cal
C}^2\otimes{\cal C}^N$  composite quantum systems }

\author{Sini\v{s}a Karnas and Maciej
Lewenstein\cite{poczta2}}
\address{Institut f\"ur Theoretische Physik,
Universit\"at Hannover, D-30167 Hannover, Germany
}
\maketitle
\date{\today}

\begin{abstract}
We investigate separability and entanglement  of mixed states in ${\cal
C}^2\otimes{\cal C}^2\otimes{\cal C}^N$  three party  quantum systems. We
show that all states with positive partial transposes that have rank $\le
N$ are separable. For the 3 qubit case ($N=2$) we prove that all states
$\rho$ that have  positive partial transposes and rank $3$ are
separable. We provide also constructive separability checks for the states  
$\rho$ that have the sum of the rank of $\rho$ and the ranks of partial
transposes with respect to all subsystems smaller than 15N-1. 
\end{abstract}

\narrowtext

\date{\today}

\section{Introduction}

In the recent years it became clear that entanglement is one of the most
important ingredients of the quantum information processing. While in the
early age of quantum mechanics, entanglement was associated with
"paradoxes" of quantum
mechanics\cite{EPR,Sch}, in the
last decade of the last century
it has been discovered that
entanglement plays an essential
role in fundamental applications of
quantum mechanics to information
processing (cf. \cite{effects,geste,Tel}). While the
characterization of separable and entangled pure states of bipartite
systems is quite well understood (cf.
\cite{peresbook}), it is not the case for mixed states. In the last four
years, however, a lot of progress has been achieved in our understanding 
of the separability and entanglement problem for bipartite systems (cf. 
\cite{primer}). The first major step was the proper definition of
separable and entangled states formulated by Werner
\cite{Werner}. The next milestone was the discovery by Peres\cite{Peres}
of the fact that all separable states are must necessarily have a positive
partial transpose
\cite{pt}. Soon after Horodeckis \cite{ho96} have shown the Peres criterium
provides also a sufficient condition for separability in two qubit
($2\times2$) and one qubit one qutrit ($2\times 3$) systems. Subsequently,
P. Horodecki \cite{tran} has constructed the first examples of the, so called, bound
entangled states, i.e. the first examples of the entangled states with
positive partial transpose (PPT ES). This discovery has stimulated great
interest in the studies of properties of  PPT ES. Some of the most
important results, in particular coming from the Horodecki family, IBM
group, and Innsbruck--Hannover collaboration are described in
Refs. \cite{primer}.

More recently, considerable interest has been devoted to 
multiparty entanglement\cite{rew2000}. The first papers on 3 qubit states
led to the discovery of the, so called, GHZ states \cite{GHZ}, which are
particularly suited to study the break down of the Bell like inequalities in
quantum mechanics \cite{peresbook}. Three party (and
multiparty, in general) entanglement of the GHZ type can allow for
interesting applications, such as for instance quantum secret sharing
\cite{sharing}, and many experimental groups have recently tried to
generate such states\cite{exper}.  

Theoretical studies of the structure of multiparty entangled states 
has just started\cite{rew2000}. 
First of all, pure state entanglement has
been investigated. An important direction of research was here initiated
by Ref. \cite{Linden}. In this paper Linden and Popescu have studied
whether a given quantum state can be transformed into another one
using local unitary (or at least non-unitary invertible)
transformations. Such a geometric approach calls for studies of
invariants of local unitary and local non-unitary invertible
transformations, and leads elegantly and naturally to the concepts of
Schmidt coefficients \cite{peresbook}, and Schmidt number\cite{terhal001}
for pure states in bipartite systems. 

This approach and concepts can be generalized to the case of 3 qubit
systems and in general for mixed states, but it is by no means an easy
task. In particular, as pointed out in Ref. \cite{Vidal}, in both
cases one expects various, locally not equivalent kinds of
entanglement to arise. Very recently the concept of Schmidt coefficients
(i.e. invariants of the local unitary transformations has been formulated
for 3 qubit systems
\cite{Acin,sudbery}. 
The other approach (based on the investigations of local non-unitary
invertible operations) has been followed by the Innsbruck group \cite{W}.
D\"ur {\it et al.} were able to show that there are essentially 3 types of
entanglement of pure states: bipartite entanglement, $W$-class
entanglement, and
$GHZ$-class entanglement.  The are ways of characterizing the 3-qubit
entanglement with the help of a, so called, tangle \cite{wootters}, and 
other local invariants \cite{Acin,sudbery}. Numerous studies of various
types of multiparty entanglement of pure states and various interesting
examples of it  have been conducted in the recent years \cite{conduct}.

At this point it is worth mentioning that while for the  
pure states in bipartite systems it is possible to quantify, 
or better to say to characterize the
entanglement in the canonical way\cite{Vidal}, 
this is not necessarily the case for
the mixed states. The studies of entanglement and
separability in mixed states of a three party system has just
begun. Among the recent results it is worth mentioning
the demonstration of separability of the states that differ not much
from fully chaotic state\cite{zyczko}, the construction of entangled
states that have all partial transposes positively defined, employing the
concept of unextendible product basis
\cite{upb}, the classification of multi-qubit states based on the
separability and distillability properties of certain partitions,  
\cite{duer},  the  generalization of the concept of mean Schmidt number
to the case of multiparty systems
\cite{jenshans}, a formulation of the necessary and sufficient conditions
for separability in term of linear maps \cite{mapy}, studies of  the
properties of relative entropy in multiparty systems 
\cite{plenio}, and studies of states symmetric with respect to trilateral unitary rotations\cite{tilo}.  
We  have presented recently\cite{ourclass} 
a classification  of mixed states for 3 qubit
systems into the  separable class,  the bipartite, the  $W$--,
and the $GHZ$--classes of states. Following the Refs.
\cite{witness} we constructed canonical form of entanglement witnesses  
for each class, and discussed their optimization.

In this paper we consider entanglement and separability 
of mixed states in ${\cal
C}^2\otimes{\cal C}^2\otimes{\cal C}^N$  three party  quantum systems. 
Such systems are of practical interests since i) for $N=2$ they
reduce to the intensively studied 3 qubit systems; ii) for $N$ large, they
can be used to describe two qubits interacting via a "bus" mode; this is
how the quantum gates can be realized in the quantum computer model based
on cold  trapped ions \cite{IO1,IO2}.

This paper generalizes the results obtained by us earlier for the case
$2\times N$ \cite{2xN} and $M\times N$ \cite{MxN} systems. We use here
the same mathematical tools that have been developed in our earlier
work \cite{M&A,pptbsa}, i.e. the method of subtracting from a
given state $\rho$ projectors on product states keeping the remainder,
as well as its partial transpose (--s) 
positively definite

The paper  is organized as follows. In section 2 we demonstrate  that all states in
${\cal C}^2\otimes{\cal C}^2\otimes{\cal C}^N$ systems with positive
partial transposes that have rank
$\le N$ are separable. This section is divided into 3 subsections, and
the main result is presented in the last subsection. In the first
subsection we present the   canonical form of the investigated states; in
the second one we prove an important Lemma that states  that for 
the 3 qubit case ($N=2$)  all states
$\rho$ that have  positive partial transposes and rank $3$ are
separable. In the section 3 we discuss constructive criteria and
separability checks for  the states  
$\rho$ that have the sum of the rank of $\rho$ and the ranks of partial
transposes with respect to all subsystems smaller equal than $15N-1$.
We discuss here the concept of the {\it edge} states, i.e. those from
which no projector on a product state can be subtracted without loosing
either the positivity, or the PPT property. We discuss here also the
methods of constructing  the, so called, entanglement witnesses,
and their canonical form.  In section 4 we specify the previous results
for the case 
$N=2$, and  we provide  constructive separability checks for the states  
$\rho$ that have the sum of the rank of $\rho$ and the ranks of partial
transposes with respect to all subsystems smaller equal than $29$.

In this paper we denote by $R(\rho)$, $K(\rho)$, $r(\rho)$ 
and $k(\rho)$ the range, 
the kernel, the rank, the dimension of the kernel of $\rho$, respectively. 
Also, $\ket{{\hat e}}$ 
will denote a vector orthogonal to $\ket{e}$. The symbol ${\rm
diag}[\sigma_1, \sigma_2, \ldots]$ denotes a matrix with diagonal blocks
$\sigma_1, \sigma_2, \ldots$.

\section{PPT states of rank $N$ in ${\cal C}^{2}\otimes{\cal C}^{2}
\otimes{\cal C}^{N}$ systems}

\subsection{Generic form of the rank $N$ PPT states}

In this section we will derive the canonical form of the separable states
in ${\cal
C}^{2}\otimes{\cal C}^{2}\otimes{\cal C}^{N}$ with $r(\rho)=N$.
The canonical form will allow for an explicit decomposition of a given 
state in terms of convex sum of projectors on product vectors.
 In the following the three  parties will be called 
Alice, Bob and Charlie. We begin with the following Lemma:

\begin{lem}\label{lemdrei1} 
{\bf :} Every PPT state $\rho$ in ${\cal C}^{2}\otimes{\cal C}^{2}\otimes{
\cal C}^{N}$ with $r(\rho)=N$, such that in some local basis
($|0_A\rangle$, $|1_A\rangle$ for Alice, $|0_B\rangle$, $|1_B\rangle$ for
Bob, $|0_C\rangle\ldots,|N-1_C\rangle$ for Charlie) without loosing the
generality we have
$r(\bra{1_{A},1_{B}}\rho\ket{1_{A},1_{B}})=N$,  can be
transformed using a  reversible local operation to the following
canonical form:
\begin{eqnarray}
\rho &=&\sqrt{D}\left(\begin{array}{cccc}
B^{\dag}C^{\dag}CB&B^{\dag}C^{\dag}C&B^{\dag}C^{\dag}B&B^{\dag}C^{\dag}\\
C^{\dag}CB&C^{\dag}C&C^{\dag}B&C^{\dag}\\
B^{\dag}CB&B^{\dag}C&B^{\dag}B&B^{\dag}\\
CB&C&B&1\end{array}\right)\sqrt{D}\nonumber\\
&=&\sqrt{D}\left(\begin{array}{c}
B^{\dag}C^{\dag}\\
C^{\dag}\\
B^{\dag}\\
1\end{array}\right)
\left(\begin{array}{cccc}
CB&C&B&1\end{array}\right)\sqrt{D},
\end{eqnarray}
where $[B,B^{\dag}]=[C,C^{\dag}]=[C,B]=[C,B^{\dag}]=0$ and $D=D^{\dag}$;
$B$,$C$ and $D$ are  operators acting in the  Charlie's space.
\end{lem}

\noindent{\bf{Proof:}} The state $\rho$ can be always written in the
considered basis as:
\begin{displaymath}
\rho=\left(\begin{array}{cccc}
E_{1}&E_{5}&E_{6}&E_{7}\\
E_{5}^{\dag}&E_{2}&E_{8}&E_{9}\\
E_{6}^{\dag}&E_{8}^{\dag}&E_{3}&E_{10}\\
E_{7}^{\dag}&E_{9}^{\dag}&E_{10}^{\dag}&E_{4}\end{array}\right),
\end{displaymath}
where  $E$'s are $N\times N$-matrices, and $r(E_4)=N$. After the
projection
$\tilde\rho=\bra{1_{A}}\rho\ket{1_{A}}$ we obtain the reduced
state
\begin{displaymath}
\tilde\rho=\left(\begin{array}{cc}
E_{3}&E_{10}\\
E_{10}^{\dag}&E_{4}\\\end{array}\right).
\end{displaymath}
After performing a reversible local non-unitary "filtering" 
$\frac{1}{\sqrt{E_{4}}}$ on Charlie's
side the matrix  $\tilde\rho$
can be written as:
\begin{displaymath}
\tilde\rho =\left(\begin{array}{cc}
A&B^{\dag}\\B& 1\\
\end{array}\right).
\end{displaymath} 
This matrix is obviously positive, i.e. can be represented as 
\cite{MxN} $\tilde \rho
=\Sigma+{\rm diag}[\Delta ,0]$, where $\Delta =A-B^{\dag}B$,  
\begin{displaymath}
\Sigma =\left(\begin{array}{cc}
B^{\dag}B&B^{\dag}\\
B& 1\\\end{array}\right).
\end{displaymath}  
The matrix $\tilde\rho$ mu{\ss} has the rank $N$. We observe that
$\Sigma$ has also the range $N$, and possesses  $N$ vectors in
its kernel
$\ket{\phi_{f}} =\ket{1}\ket{f}-\ket{2} B\ket{f}$. We will show
that 
$\Delta =0$.

Using the fact that $\tilde\rho\geq 0$, we observe that
$\Delta\geq 0$. But, since $r(\tilde\rho)=r(\Sigma)$, the ranges of the
the matrices must fulfill
$R(\tilde\rho)=R(\Sigma)\supseteq R({\rm diag}[\Delta,0])$, so that the
corresponding kernels fulfill
$K({\rm diag}[\Delta,0])\supseteq K(\Sigma)$. The kernel  $K(\Sigma)$ is
spanned by the vectors of the form 
$\ket{\phi_{f}}=\ket{1}\ket{f}-\ket{2}B\ket{f}$, where $\ket{f}$ is
arbitrary,  for which  $\bra{\phi_{f}}{\rm
diag}[\Delta,0]\ket{\phi_{f}}=0$ must hold also. This means, however, 
that $\Delta\ket{f}=0$ for all
$\ket{f}$, and thus $\Delta=0$.

The fact that $B$ is a normal operator follows from the fact that
$\tilde\rho^{t_{A}}$ must be positively definite. This condition
implies that $BB^{\dag}-B^{\dag}B\geq 0$. The latter positive
operator has, however, the trace zero, and must therefore vanish, i.e.
$[B,B^{\dag}]=0$.

Similarly, if we consider the projection $\bra{1_{B}}\rho\ket{1_{B}}$,
for the same reasons as above we conclude that the resulting matrix  
\begin{displaymath}
{\bar \rho}=\left(\begin{array}{cc}
C^{\dag}C^{\dag}&C^{\dag}\\
C& 1\\\end{array}\right),
\end{displaymath} 
with $[C,C^{\dag}]=0$. Summarizing, after performing a local filtering
operation 
$\frac{1}{\sqrt{E_{4}}}$ we can bring the matrix $\rho$ to the form:
\begin{displaymath}
{\rho}=\left(\begin{array}{cccc}
E_{1}&E_{5}&E_{6}&E_{7}\\
E_{5}^{\dag}&C^{\dag}C&E_{8}&C^{\dag}\\
E_{6}^{\dag}&E_{8}^{\dag}&B^{\dag}B&B\\
E_{7}^{\dag}&C^{\dag}&B& 1\end{array}\right).
\end{displaymath}

Now,  the matrix $\rho$ possesses as kernel vectors
$\ket{10}\ket{f}-\ket{11} B\ket{f}$ and  
$\ket{01}\ket{g}-\ket{11} C\ket{g}$ for all $\ket{f},\ket{g}$ from the
Charlie's space.  This implies that we must have 
$E_{8}=C^{\dag}B$,
$E_{6}=E_{7}B$ and $E_{5}=E_{7}C$. The matrix $\rho$ has thus the form:
\begin{displaymath}
\rho=\left(\begin{array}{cccc}
E_{1}&E_{7}C&E_{7}B&E_{7}\\
C^{\dag}E_{7}^{\dag}&C^{\dag}C&C^{\dag}C&C^{\dag}\\
B^{\dag}E_{7}^{\dag}&B^{\dag}C&B^{\dag}B&B^{\dag}\\
E_{7}^{\dag}&C&B& 1\end{array}\right).
\end{displaymath}
In the next step we consider its partial transpose with respect to Alice
given by:
\begin{displaymath}
\rho^{t_{A}}=\left(\begin{array}{cccc}
E_{1}&E_{7}C&B^{\dag}E_{7}^{\dag}&B^{\dag}C\\
C^{\dag}E_{7}^{\dag}&C^{\dag}C&E_{7}^{\dag}&C^{\dag}\\
E_{7}B&E_{7}&B^{\dag}B&B^{\dag}\\
C^{\dag}B&C^{\dag}&B& 1\end{array}\right).
\end{displaymath}
Since partial transpose with respect to Alice is positive and does not
change 
$\bra{1_{A}}\rho\ket{1_{A}}$, the  vectors  
$\ket{10}\ket{f}-\ket{11} B\ket{f}$ should remain in the kernel. This
implies the equality
$E_{7}=B^{\dag}C^{\dag}$, and the following form of 
$\rho$:
\begin{displaymath}
\rho=\left(\begin{array}{cccc}
E_{1}&B^{\dag}C^{\dag}C&B^{\dag}C^{\dag}B&B^{\dag}C^{\dag}\\
C^{\dag}CB&C^{\dag}C&C^{\dag}B&C^{\dag}\\
B^{\dag}CB&B^{\dag}C&B^{\dag}B&B^{\dag}\\
CB&C&B& 1\end{array}\right).
\end{displaymath}
The above form can be rewritten as 
\begin{displaymath}
\rho=\left(\begin{array}{cccc}
B^{\dag}C^{\dag}CB&B^{\dag}C^{\dag}C&B^{\dag}C^{\dag}B&B^{\dag}C^{\dag}\\
C^{\dag}CB&C^{\dag}C&C^{\dag}B&C^{\dag}\\
B^{\dag}CB&B^{\dag}C&B^{\dag}B&B^{\dag}\\
CB&C&B& 1\end{array}\right)+{\rm diag}[{\tilde\Delta},0,0,0],
\end{displaymath}
where $\Delta=E_{1}-B^{\dag}C^{\dag}CB$. Using the short hand notation
we get
\begin{displaymath}
\rho=\left(\begin{array}{c}
B^{\dag}C^{\dag}\\
C^{\dag}\\
B^{\dag}\\
1\end{array}\right)
\left(\begin{array}{cccc}
CB&C&B&1\end{array}\right)+{\rm diag}[{\tilde\Delta},0,0,0].
\end{displaymath}
The first term in  $\rho$ is PPT and has the following 
$3N$ vectors in the kernel: 
\begin{eqnarray}
\ket{\psi}&=&\ket{00}\ket{f}+\ket{11}CB\ket{f}\nonumber\\
\ket{\phi}&=&\ket{01}\ket{g}+\ket{11}C\ket{g}\nonumber\\
\ket{\chi}&=&\ket{10}\ket{h}+\ket{11}B\ket{h}\nonumber,
\end{eqnarray}
for arbitrary $\ket{f},\ket{g}$ and $\ket{h}$. Similarly as above, this
means that as in the case of $\Delta$, the matrix 
$\tilde\Delta$ must vanish. This provides us with the final form of 
$\rho$:
\begin{eqnarray}
\rho &=&\left(\begin{array}{cccc}
B^{\dag}C^{\dag}CB&B^{\dag}C^{\dag}C&B^{\dag}C^{\dag}B&B^{\dag}C^{\dag}\\
C^{\dag}CB&C^{\dag}C&C^{\dag}B&C^{\dag}\\
B^{\dag}CB&B^{\dag}C&B^{\dag}B&B^{\dag}\\
CB&C&B&1\end{array}\right)\\
&=&\left(\begin{array}{c}
B^{\dag}C^{\dag}\\
C^{\dag}\\
B^{\dag}\\
1\end{array}\right)
\left(\begin{array}{cccc}
CB&C&B&1\end{array}\right).
\end{eqnarray}
It remains only to prove the commutation relations 
$[B,C]=[B,C^{\dag}]=0$. This follows from the positivity of all partial
transposes of
$\rho$. In particular, $\rho^{t_{A}}$ is:
\begin{eqnarray}
\rho^{t_{A}}=\left(\begin{array}{c}
B^{\dag}C\\
C\\
B^{\dag}\\
1\end{array}\right)
\left(\begin{array}{cccc}
C^{\dag}B&C^{\dag}&B&1\end{array}\right),
\end{eqnarray}
which is obviously positive definite.

In contrast, $\rho^{t_{B}}$ can be written as:
\begin{eqnarray}
\rho^{t_{B}}=\left(\begin{array}{cccc}
B^{\dag}C^{\dag}CB&C^{\dag}CB&B^{\dag}C^{\dag}B&C^{\dag}B\\
B^{\dag}C^{\dag}C&C^{\dag}C&B^{\dag}C^{\dag}&C^{\dag}\\
B^{\dag}CB&CB&B^{\dag}B&B\\
B^{\dag}C&C&B^{\dag}& 1\end{array}\right).
\end{eqnarray}
Because of its positivity, the matrix $\rho^{t_{B}}$ must  possess the
kernel vector
$\ket{01}\ket{g}-\ket{11}C\ket{g}$, which implies that $[C,B]=0$.
The matrix $\rho^{t_{B}}$ can be then written as:
\begin{eqnarray}
\rho^{t_{B}}=\left(\begin{array}{c}
C^{\dag}B\\
C^{\dag}\\
B\\
1\end{array}\right)
\left(\begin{array}{cccc}
B^{\dag}C&C&B^{\dag}&1\end{array}\right),
\end{eqnarray} 
which implies automatically the positivity. 
It remains finally to consider 
$\rho^{t_{AB}}$. The latter can be written as:
\begin{eqnarray}
\rho^{t_{AB}}=\left(\begin{array}{cccc}
B^{\dag}C^{\dag}CB&C^{\dag}CB&B^{\dag}CB&CB\\
B^{\dag}C^{\dag}C&C^{\dag}C&B^{\dag}C&C\\
B^{\dag}C^{\dag}B&C^{\dag}B&B^{\dag}B&B\\
B^{\dag}C^{\dag}&C^{\dag}&B^{\dag}& 1\end{array}\right).
\end{eqnarray}
From the positivity of  
$\rho^{t_{AB}}$ follows that $\ket{10}-\ket{11}B^{\dag}\ket{f}$
is a kernel vector, so that
$[B^{\dag},C]=0$ must hold. This in turn allows to write $\rho^{t_{AB}}$
as:
\begin{eqnarray}
\rho^{t_{AB}}=\left(\begin{array}{c}
CB\\
C\\
B\\
1\end{array}\right)
\left(\begin{array}{cccc}
B^{\dag}C^{\dag}&C^{\dag}&B^{\dag}&1\end{array}\right).
\end{eqnarray}
Again, this form assures positive definiteness, and concludes the proof of
the  Lemma. ${\Box}$

Now, we are in the position to prove:
\begin{lem}\label{seplem}{\bf :}
A PPT--state $\rho$ in $\quad{\cal C}^{2}\otimes{\cal
C}^{2}\otimes{\cal C}^{N}$, whose rank  $r(\rho)=N$, and for
which there exists a product basis
$\ket{e_{A},f_{B}}$, such  that
$r(\bra{e_{A},f_{B}}\rho\ket{e_{A},f_{B}})=N$, is separable.
\end{lem}
{\bf{Proof:}} 
The state $\rho$ can be written according to the Lemma (\ref{lemdrei1})
as
\begin{eqnarray}
\rho=\left(\begin{array}{c}
B^{\dag}C^{\dag}\\
C^{\dag}\\
B^{\dag}\\
1\end{array}\right)
\left(\begin{array}{cccc}
CB&C&B&1\end{array}\right).
\end{eqnarray}
Since all operators commute, they have to have 
common  eigenvectors $\ket{f_{n}}$, with eigenvalues $b_n,c_n$,
respectively, and 
\begin{eqnarray}
\bra{f_{n}}\rho\ket{f_{n}}=\left(\begin{array}{c}
b_{n}^{*}c_{n}^{*}\\
c_{n}^{*}\\
b_{n}^{*}\\
1\end{array}\right)
\left(\begin{array}{cccc}
c_{n}b_{n}&c_{n}&b_{n}&1\end{array}\right).
\end{eqnarray}
This is, however, a product vector in Alice's and Bob's spaces.
We can thus write  $\rho$  as
$\rho=\sum_{n=1}^{N}\proj{\psi_{n}}\otimes\proj{\phi_{n}}
\otimes\proj{f_{n}}$.
Because the local transformations  used above were reversible, we can
now apply their inverses and obtain a decomposition of the initial state
$\rho$ in a sum of projectors onto product vectors. This proves
separability of $\rho$, and the Lemma.$\Box$

From the Lemma 1 and 2 we conclude that in order to prove that PPT states
$\rho$  supported on ${\cal C}^{2}\otimes{\cal C}^{2}\otimes{\cal C}^{N}$
with
$r(\rho)=N$   are separable, it is enough to show that one can find a
product basis such that 
$r(\bra{e_{A},f_{B}}\rho\ket{e_{A},f_{B}})=N$. We will accomplish the
proof in another way. Instead, we will prove the separability directly,
and the desired canonical form of $\rho$ will be a consequence of that.
To this aim we will use the results of Ref.
\cite{2xN}, and the following theorem from Ref.
\cite{MxN}.
\begin{theo}\label{prodmn}{\bf : }
For all PPT states  $\rho$ that are supported on $M\times N$--space
$(M\leq N)$, and that have rank $N$, there exists  a product basis
such that 
$r(\bra{1_{A}}\rho\ket{1_{A}})=N$ and $\rho$ is separable and has the
form 
\begin{equation}
\rho=\sum_{i=1}^{N}\proj{e_{i},b_{i}}
\end{equation}
where  $\ket{b_{i}}$ are linearly  independent.  Additionally, the above
decomposition is unique.  
\end{theo}

This theorem can be used to prove the following Lemma:
\begin{lem}{\bf : }
Any PPT state $\rho$ supported on  ${\cal C}^{2}\otimes{\cal
C}^{2}\otimes{\cal C}^{N}$ with $N\geq 4$, for which   $r(\rho)=N$, is
separable, and obeys assumptions of Lemma $1$.
\end{lem}
{\bf{Proof:}} A ${\cal C}^{2}\otimes{\cal C}^{2}\otimes{\cal
C}^{N}$--system can be regarded as a ${\cal C}^{4}\otimes{\cal
C}^{N}$--system. From the theorem 
\ref{prodmn} we obtain that:
\begin{equation}
\rho=\proj{\psi_{AB_{1}}}\otimes\proj{C_{1}}
+\sum_{i=2}^{N}\proj{\psi_{AB_{i}},C_{i}}.\label{suma}
\end{equation}
Note, however, that we can find now a vector  
$\ket{C}$ in Charlie's space, so that
$\bra{C}\rho\ket{C}\sim\proj{\psi_{AB_{1}}}$. Because the state
$\rho$ has a PPT property with respect to all partitions,
$\proj{\psi_{AB_{1}}}$  must be PPT with respect to Alice or Bob, 
i.e. it must be a product
state. This observation concerns all projectors that enter the convex
sum (\ref{suma}), so that we conclude that  $\rho$ is  separable. From 
($12$) it follows directly that $\rho$ can be projected onto
$\ket{1_{A},1_{B}}$, so that
$r(\bra{1_{A},1_{B}}\rho\ket{1_{A},1_{B}}=N$.${\Box}$

\subsection{Separability of states with rank $\le 3$ in ${\cal C}^{2}\otimes{\cal C}^{2}\otimes{\cal C}^{2}$}

Now we have to consider the cases $N=2,3$. The following Corollary and Lemma deal with the case 
$N=2$:

\begin{kor}{\bf :}
Any  PPT state $\rho$ supported on  
${\cal C}^{2}\otimes{\cal C}^{2}\otimes{\cal C}^{2}$ and such that  $r(\rho)=2$, has a product vector 
$\ket{e,f,g}$ in its kernel.
\end{kor}
{\bf{Proof:}} The vector $\ket{e,f,g}$ belongs to the kernel iff it is orthogonal to the range, i.e. iff it
is orthogonal to the two vectors
$\{\ket{\psi_{1}},\ket{\psi_{2}}\}$, which span the range of $\rho$. We
can choose 
$\ket{e}$ arbitrary and set  $\ket{f}=\ket{0}+\alpha\ket{1}$, 
so that we obtain two equations:
\begin{equation}
(\braket{\psi_{i}}{e,0}+\alpha\braket{\psi_{i}}{e,1})\ket{g}=0
\end{equation}
We treat these equations as linear homogeneous equations for $|g\rangle$; they have nontrivial solutions
if the corresponding determinant of a $2\times 2$ matrix vanishes. 
This gives a quadratic 
equation for 
$\alpha$, which has always at least one solution. ${\Box}$

\begin{lem}{\bf : }
Any PPT state $\rho$, supported on  ${\cal C}^{2}\otimes{\cal C}^{2}\otimes{\cal C}^{2}$,
and such that  $r(\rho)=2$, is separable (compare \cite{smolin}).
\end{lem}
{\bf{Proof:}}
If $\ket{e_{A},f_{B},g_{C}}$ is in the kernel of $\rho$, then PPT property implies that also
$\rho^{t_{A}}\ket{e^{*}_{A},f_{B},g_{C}}=0$. We obtain then that
$\bra{\hat{e}^{*}_{A}}\rho^{t_{A}}\ket{e^{*}_{A},f_{B},g_{C}}=0$, where $\ket{\hat{e}_{A}}$ is orthogonal to $\ket{e_{A}}$.    
 This equation is equivalent to $\bra{e_{A}}\rho\ket{\hat{e}_{A},f_{B},g_{C}}=0$. This, however, means that
$\rho\ket{\hat{e}_{A},f_{B},g_{C}}=\ket{\hat{e}_{A},\psi_{BC}}$, where
$\ket{\psi_{BC}}$ a vector  in Bob's and Charlie's space. 
Now, according to the Lemma  $2$ of Ref. \cite{2xN} which deals with ${\cal
  C}^{2}\otimes{\cal C}^{N}$--systems, we can subtract the projector  $\proj{\hat{e}_{A},\psi_{BC}}$
from $\rho$, so that
$\tilde{\rho}=\rho-\frac{1}{\bra{\hat{e}_{A},\psi_{BC}}\rho^{-1}\ket{\hat{e}_{A},\psi_{BC}}}\proj{\hat{e}_{A},\psi_{BC}}$
is positive, has rank 1, i.e. is a projector. Since it has the PPT property with respect to Alice's
system, it must be separable with respect to $A-BC$ partition. 
In general, we can thus write
$\rho=\tilde{\Lambda}\proj{\tilde{e}_{A}}\otimes\proj{\tilde{\psi}_{BC}}+\Lambda\proj{\hat{e}_{A},\psi_{BC}}$.
Projecting onto $\ket{e_{A}}$ we get  
$\bra{e_{A}}\rho\ket{e_{A}}\propto\proj{\tilde{\psi}_{BC}}$. Since 
$\rho$  has the PPT property with respect to all partitions, the
projector $\proj{\tilde{\psi}_{BC}}$  must project onto a product
vector.  The same can be of course said  about $\proj{\psi_{BC}}$, since
the projection onto  $\proj{\hat{\tilde{e}}_{A}}$ gives 
$\bra{\hat{\tilde{e}}_{A}}\rho\ket{\hat{\tilde{e}}_{A}}\propto
\proj{\psi_{BC}}$,
which implies that
$\proj{\psi_{BC}}$ is a product state  and concludes the proof.
${\Box}$

Now we have still to prove the case  $N=3$. Before we do that, however, we need one more Corollary and Lemma 
concerning the case $N=2$:
\begin{kor}{\bf : }
Any PPT state $\rho$, supported on ${\cal C}^{2}\otimes{\cal C}^{2}\otimes{\cal C}^{2}$, such that
$r(\rho)=3$, has a product vector  $\ket{e,f,g}$ in the kernel.
\end{kor}
{\bf{Proof:}}
Let  $\rho$ be  PPT--state in ${\cal C}^{2}\otimes{\cal C}^{2}\otimes{\cal C}^{2}$. It can be regarded 
as a 
${\cal C}^{2}_{A}\otimes{\cal C}^{4}_{BC}$--state. According to Theorem
$1$ of Ref. \cite{2xN} this state is supported on  ${\cal C}^{2}_{A}\otimes{\cal C}^{3}_{BC}$,
and must have the form:
\begin{equation}
\rho=\sum_{i=1}^{3}\proj{e_{A_{i}}}\otimes\proj{\psi_{BC_{i}}}.
\end{equation}
We take  $\ket{e}$ orthogonal to $\ket{e_{A_{3}}}$, and demand that $\ket{f,g}$ is
orthogonal to $\ket{\psi_{BC_{1}}}$ and $\ket{\psi_{BC_{2}}}$. Setting
$\ket{f_{B}}=\ket{0}_{B}+\alpha\ket{1}_{B}$, we obtain the following system of linear homogeneous  equations
for $\ket{g}$:
\begin{equation}
(\braket{\psi_{BC_{i}}}{0_{B}}+\alpha\braket{\psi_{BC_{i}}}{1_{B}})\ket{g}=0
\end{equation}
for $i=1,2$.  These equations posses a nontrivial solution if the corresponding determinant of the
2$\times2$ matrix vanish. This lead to a quadratic equation for $\alpha$, which has always a
solution, and that proves the Corollary.${\Box}$
 
The existence of product vectors in the kernel is used in the proof of the Lemma below. This Lemma provides
one of the most important results of this paper: it implies that in ${\cal C}^{2}\otimes{\cal
C}^{2}\otimes{\cal C}^{2}$ systems there is no PPT entanglement of rank smaller than 4. 

\begin{lem}\label{zweidrei}{\bf :}
Any PPT state $\rho$, supported on  ${\cal C}^{2}\otimes{\cal C}^{2}\otimes{\cal C}^{2}$, such that
$r(\rho)=3$, is separable.
\end{lem}   
{\bf{Proof:}}
Let $|e_A, f_B,g_C\rangle$ belongs to $K(\rho)$. From the condition
$\rho\ket{e_{A},f_{B},g_{C}}=0$ for the product vector in the kernel,
follows that:
\begin{eqnarray}
\bra{e_{A}}\rho\ket{\hat{e}_{A},f_{B},g_{C}}&=&0,\nonumber\\
\bra{f_{B}}\rho\ket{e_{A},\hat{f}_{B},g_{C}}&=&0,\nonumber\\
\bra{g_{C}}\rho\ket{e_{A},f_{B},\hat{g}_{C}}&=&0.\nonumber
\end{eqnarray}
This means that
\begin{eqnarray}
\rho\ket{\hat{e}_{A},f_{B},g_{C}}&=&\ket{\hat{e}_{A}}\ket{\psi_{BC}},\nonumber\\
\rho\ket{e_{A},\hat{f}_{B},g_{C}}&=&\ket{\hat{f}_{B}}\ket{\psi_{AC}},\nonumber\\
\rho\ket{e_{A},f_{B},\hat{g}_{C}}&=&\ket{\hat{g}_{C}}\ket{\psi_{AB}}.\nonumber
\end{eqnarray}
We define 
\begin{equation}
\tilde\rho=\rho-\lambda\proj{\hat{e}_{A}}\otimes\proj{\psi_{BC}},
\end{equation}
where $\lambda=\frac{1}{\bra{\hat{e}_{A},\psi_{BC}}\rho^{-1}\ket{\hat{e}_{A},\psi_{BC}}}$ (see Lemma $2$
of Ref. \cite{2xN}). Now, $\tilde\rho$ is a  PPT state with respect to $A-BC$ partition, i.e.
$\tilde{\rho}^{t_{A}}\ge 0$; this state has the rank 
$r(\tilde\rho)=2$. We rewrite $\tilde\rho$ as:
\begin{equation}
\tilde\rho=\lambda_{1}\proj{\hat{f}_{B}}\otimes\proj{\psi_{AC}}+\lambda_{2}\proj{\hat{g}_{C}}
\otimes\proj{\psi_{AB}}.
\end{equation}
We redefine now $\ket{e_{A}}=\ket{0}$ and $\ket{\hat{e}_{A}}=\ket{1}$, i.e. change 
the basis in Alice's
system, and represent the vectors $\ket{\psi_{AC}}$ and $\ket{\psi_{AB}}$ in the new basis as:
\begin{eqnarray}
\ket{\psi_{AC}}&=&\ket{0}\ket{\psi_{C}^{1}}+\ket{1}\ket{\psi_{C}^{2}},\\
\ket{\psi_{AB}}&=&\ket{0}\ket{\phi_{B}^{1}}+\ket{1}\ket{\phi_{B}^{2}}.
\end{eqnarray} 
In the matrix form  $\tilde\rho$ can be written as:
\begin{displaymath}
\scriptsize{\tilde\rho=
\left(\begin{array}{cc}
\left(\begin{array}{c}\lambda_{1}\proj{\hat{f}_{B}}\otimes\proj{\psi_{C}^{1}}\\
+\lambda_{2}\proj{\phi_{B}^{1}}\otimes\proj{\hat{g}_{C}}\end{array}\right)
&\left(\begin{array}{c}\lambda_{1}\proj{\hat{f}_{B}}\otimes\ket{\psi_{C}^{1}}\bra{\psi_{C}^{2}}\\
+\lambda_{2}\ket{\phi_{B}^{1}}\bra{\phi_{B}^{2}}\otimes\proj{\hat{g}_{C}}\end{array}\right)\\
\left(\begin{array}{c}\lambda_{1}\proj{\hat{f}_{B}}\otimes\ket{\psi_{C}^{2}}\bra{\psi_{C}^{1}}\\
+\lambda_{2}\ket{\phi_{B}^{2}}\bra{\phi_{B}^{1}}\otimes\proj{\hat{g}_{C}}\end{array}\right)
&\left(\begin{array}{c}\lambda_{1}\proj{\hat{f}_{B}}\otimes\proj{\psi_{C}^{2}}\\
+\lambda_{2}\proj{\phi_{B}^{2}}\otimes\proj{\hat{g}_{C}}\end{array}\right)
\end{array}\right)}.
\end{displaymath}
From the positivity of  $\tilde\rho$ and $\tilde{\rho}^{t_{A}}$ follows that when a diagonal block 
$\left(\begin{array}{c}\lambda_{1}\proj{\hat{f}_{B}}\otimes\proj{\psi_{C}^{2}}\\
+\lambda_{2}\proj{\phi_{B}^{2}}\otimes\proj{\hat{g}_{C}}\end{array}\right)$ 
acting on   $\ket{\hat{\phi}_{B}^{2}\hat{\psi}_{C}^{2}}$ vanishes, the same must be true for the 
off--diagonal block in left lower corner. Similarly, the same observation concerns the diagonal 
$\left(\begin{array}{c}\lambda_{1}\proj{\hat{f}_{B}}\otimes\proj{\psi_{C}^{1}}\\
+\lambda_{2}\proj{\phi_{B}^{1}}\otimes\proj{\hat{g}_{C}}\end{array}\right)$, the vector
$\ket{\hat{\phi}_{B}^{1}\hat{\psi}_{C}^{1}}$, and the off--diagonal block in the right upper corner. This
leads to the system of equations:
\begin{eqnarray}
\braket{\hat{f}_{B}}{\hat{\phi}_{B}^{1}}\braket{\psi_{C}^{2}}{\hat{\psi}_{C}^{1}}&=&0,\\
\braket{\phi_{B}^{2}}{\hat{\phi}_{B}^{1}}\braket{\hat{g_{C}}}{\hat{\psi}_{C}^{1}}&=&0,\\
\braket{\hat{f}_{B}}{\hat{\phi}_{B}^{2}}\braket{\psi_{C}^{1}}{\hat{\psi}_{C}^{2}}&=&0,\\
\braket{\phi_{B}^{1}}{\hat{\phi}_{B}^{2}}\braket{\hat{g_{C}}}{\hat{\psi}_{C}^{2}}&=&0.
\end{eqnarray}
This system of equations implies that at least one of the  projectors  $\proj{\psi_{AB}}$ and
$\proj{\psi_{AC}}$ must be a product state.  If it is, for instance, 
$\proj{\psi_{AB}}$, then  $\ket{\phi_B^1}=\ket{\phi_B^2}=\ket{\hat{f}_B}$ and $\rho$ becomes 
\begin{eqnarray}
\rho&=&\lambda_{1}\proj{\hat{f}_{B}}\otimes\proj{\psi_{AC}}\nonumber\\
&+&\lambda_{2}\proj{\tilde{e}_{A}}\otimes\proj{\hat{f}_{B}}\otimes\proj{\hat{g}_{C}}\nonumber\\
&+&\lambda\proj{\hat{e}}\otimes\proj{\psi_{BC}}\nonumber\\
&=&\proj{\hat{f}_{B}}\otimes\underbrace{(\lambda_{1}\proj{\psi_{AC}}+\lambda_{2}\proj{\tilde{e}_{A}}\otimes\proj{\hat{g}_{C}})}_{\sigma}\nonumber\\
&+&\lambda\proj{\hat{e}}\otimes\proj{\psi_{BC}}.
\end{eqnarray}
The operator $\sigma$ is a PPT state of rank 2 in  ${\cal C}^{2}\otimes{\cal C}^{2}$ space of Alice and
Charlie. From Peres-Horodecki criterium \cite{Peres,ho96} follows that it is separable. The matrix
$\rho$ can thus be written as 
\begin{eqnarray}
\rho&=&\lambda_{1}\proj{\bar{e}_{A}}\otimes\proj{\hat{f}_{B}}\otimes\proj{\tilde{g}_{C}}\nonumber\\
&+&\lambda_{2}\proj{\tilde{e}_{A}}\otimes\proj{\hat{f}_{B}}\otimes\proj{\hat{g}_{C}}\nonumber\\
&+&\lambda\proj{\hat{e}}\otimes\proj{\psi_{BC}}.\nonumber
\end{eqnarray}
For the above proof Alice is in no way distinguished. We can also write
\begin{eqnarray}
\bar{\rho}&=&\rho-\bar{\lambda}\proj{\tilde{e}_{A}}\otimes\proj{\hat{f}_{B}}\otimes\proj{\hat{g}_{C}}
\nonumber\\
&=&\lambda_{1}\proj{\bar{e}_{A}}\otimes\proj{\hat{f}_{B}}\otimes\proj{\tilde{g}_{C}}\nonumber\\
&+&\lambda\proj{\hat{e}_{A}}\otimes\proj{\psi_{BC}},\nonumber
\end{eqnarray} 
where 
$\bar{\lambda}\equiv\lambda_{2}
=\frac{1}{\bra{\tilde{e}_{A}\hat{f}_{B}\hat{g}_{C}}\rho^{-1}\ket{\tilde{e}_{A}\hat{f}_{B}\hat{g}_{C}}}$, 
and $\bar{\rho}$ is a 
 PPT state with respect to $C-AB$ partition. The projection of  $\bar{\rho}$ onto
$\ket{\hat{\tilde{e}}_{A}}$ gives $\bra{\hat{\tilde{e}}_{A}}\bar{\rho}\ket{\hat{\tilde{e}}_{A}}\sim\proj{\psi_{BC}}$.
This means, however, that  $\ket{\psi_{BC}}$ must be a product vector, and that concludes the
proof. ${\Box}$

\subsection{Separability of states of rank $N$ in ${\cal C}^{2}\otimes{\cal C}^{2}\otimes{\cal C}^{N}$
systems}

Now we are in the position to prove the main theorem of this section. Before that we have to complete,
however, the discussion of the case $N=3$. To this aim we prove the following Lemma:
\begin{lem}{\bf :}
Any PPT state $\rho$, supported on ${\cal C}^{2}\otimes{\cal C}^{2}\otimes{\cal C}^{3}$, such that
$r(\rho)=3$, is separable.
\end{lem}
{\bf{Proof:}} We consider the system  ${\cal C}^{2}\otimes{\cal C}^{2}\otimes{\cal C}^{3}$-System as a
${\cal C}^{4}_{AB}\otimes{\cal C}_{C}^{3}$ system. According to the  Theorem (\ref{prodmn}) 
three possibilities may occur:
\begin{itemize}
\item The state is supported on ${\cal C}^{3}_{AB}\otimes{\cal C}_{C}^{3}$. Then the density matrix must
have a form 
\begin{eqnarray}
\rho&=&\Lambda_{1}\proj{e_{AB_{1}}}\otimes\proj{f_{C_{1}}}\nonumber\\
&+&\Lambda_{2}\proj{e_{AB_{2}}}\otimes\proj{f_{C_{2}}}\nonumber\\
&+&\Lambda_{3}\proj{e_{AB_{3}}}\otimes\proj{f_{C_{3}}}\nonumber.
\end{eqnarray}
Since the vectors $\ket{f_{C_{i}}}$ are linearly independent, we can find a vector  $\ket{C}$ in Charlie's
system  such that $\bra{C}\rho\ket{C}\sim\proj{e_{AB_{1}}}$. Because the considered state has the PPT
property with respect to  all partition, the projected state $\proj{e_{AB_{1}}}$ is also PPT, and as such
must be a product state.
\item The state  is supported on ${\cal C}^{2}_{AB}\otimes{\cal C}_{C}^{3}$. The same method of
 projecting
onto appropriately chosen vector in Charlie's space allows to prove the separability. 
\item The state is supported  ${\cal C}^{3}_{AB}\otimes{\cal C}_{C}^{2}$. That is, however, 
nothing else
but a state  in ${\cal C}^{2}\otimes{\cal C}^{2}\otimes{\cal C}^{2}$ system with rank 3. 
Its separability
follows from Lemma 5.
\end{itemize}
This concludes the proof of the Lemma 6.${\Box}$

Now, all the above presented results can be brought together in a form of the following theorem:
\begin{theo}{\bf : }
Every PPT state, supported on  ${\cal C}^{2}\otimes{\cal C}^{2}\otimes{\cal C}^{N}$, such that
$r(\rho)=N$, is separable and has the canonical form of the Lemma \ref{lemdrei1},
\begin{eqnarray}
\rho &=&\sqrt{D}\left(\begin{array}{cccc}
B^{\dag}C^{\dag}CB&B^{\dag}C^{\dag}C&B^{\dag}C^{\dag}B&B^{\dag}C^{\dag}\\
C^{\dag}CB&C^{\dag}C&C^{\dag}B&C^{\dag}\\
B^{\dag}CB&B^{\dag}C&B^{\dag}B&B^{\dag}\\
CB&C&B&1\end{array}\right)\sqrt{D}\\
&=&\sqrt{D}\left(\begin{array}{c}
B^{\dag}C^{\dag}\\
C^{\dag}\\
B^{\dag}\\
1\end{array}\right)
\left(\begin{array}{cccc}
CB&C&B&1\end{array}\right)\sqrt{D},
\end{eqnarray}
 where  $B$,$C$ and $D$ are
operators acting in Charlie's space that fulfill $[B,B^{\dag}]=[C,C^{\dag}]=[C,B]=
[C,B^{\dag}]=0$ and
$D=D^{\dag}$.
\end{theo}

In the next section we will study states in ${\cal C}^{2}\otimes{\cal C}^{2}\otimes{\cal C}^{N}$ 
with
low ranges, but $\ge N$. By looking at product vectors in the ranges of $\rho$ and its partial 
transposes
it is posiible to check separability for low rank matrices, similarly as in the case of bipartite 
systems 
in ${\cal C}^{M}\otimes{\cal C}^{N}$  \cite{2xN,MxN}.

\section{Separability checks and criteria  for generic low rank states in $\quad{\cal
C}^2\otimes{\cal C}^2\otimes{\cal C}^N$ systems}

In this section we will study the 
 PPT states $\rho$ that 
posses a finite number of product vectors in their range 
$\ket{e_{i},f_{i},g_{i}}\in{\cal
  C}^2\otimes{\cal C}^2\otimes{\cal C}^N$ such that
$\ket{e_{A_{i}},f_{B_{i}},g_{C_{i}}}\in R(\rho)$, $\ket{e_{A_{i}}^{*},f_{B_{i}},g_{C_{i}}}\in
R(\rho^{t_{A}})$, $\ket{e_{A_{i}},f_{B_{i}}^{*},g_{C_{i}}}\in R(\rho^{t_{B}})$ and
$\ket{e_{A_{i}}^{*},f_{B_{i}}^{*},g_{C_{i}}}\in R(\rho^{t_{AB}})$. We will show that this is 
generically the case when 
$r(\rho)+r(\rho^{t_{A}})+r(\rho^{t_{B}})+r(\rho^{t_{AB}})\leq 15N-1$. Let us call the
 set of such vectors
$V[\rho]$ The search for the desired 
product vectors $\{\ket{e_{A_{i}},f_{B_{i}},g_{C_{i}}}\}\in{\cal C}^2\otimes{\cal
  C}^2\otimes{\cal C}^N$ is reduced to the problem of solving a system of 
multipolynomial equations \cite{2xN}. When the number of equations is equal to (bigger than) 
the number of
available unknown parameters, one expect the number of solutions to be  finite (zero). 
The states of low
ranks fulfilling this property has been termed {\it {generic}} in Ref. \cite{MxN}. In particular, the
states for which the number of the desired vectors in any of the considered ranges is smaller that
the corresponding rank, must be entangled. Particularly important are states that do not contain
any product vector of the above described properties in the range. Such states are termed {\it edge states},
and play major role in characterization and classification of the PPT entangled states \cite{witness}.
 
\subsection{Generic states}
Let $\ket{K_{i}}$, $\ket{K_{A_{i}}}$, $\ket{K_{B_{i}}}$ and
$\ket{K_{AB_{i}}}$ are linearly independent vector that span  the kernels of $\rho$,
$\rho^{t_{A}}$, $\rho^{t_{B}}$ and $\rho^{t_{AB}}$, respectively, so that:
\begin{eqnarray}
K(\rho)&=&{\rm{span}}\{\ket{K_{i}},i=1,\dots,k(\rho)\},\nonumber\\
K(\rho^{t_{A}})&=&{\rm{span}}\{\ket{K_{A_{i}}},i=1,\dots,k(\rho^{t_{A}})\},\nonumber\\
K(\rho^{t_{B}})&=&{\rm{span}}\{\ket{K_{B_{i}}},i=1,\dots,k(\rho^{t_{B}})\},\nonumber\\
K(\rho^{t_{AB}})&=&{\rm{span}}\{\ket{K_{AB_{i}}},i=1,\dots,k(\rho^{t_{AB}})\}.\nonumber
\end{eqnarray}
Choosing an orthonormal basis in Alice's and Bob's space we can write those vectors as:
\begin{eqnarray}
\ket{K_{i}}&=&\ket{00}\ket{k_{i}^{00}}+\ket{01}\ket{k_{i}^{01}}+\ket{10}\ket{k_{i}^{10}}+\ket{11}\ket{k_{i}^{11}}\nonumber\\
\ket{K_{A_{i}}}&=&\ket{00}\ket{k_{A_{i}}^{00}}+\ket{01}\ket{k_{A_{i}}^{01}}+\ket{10}\ket{k_{A_{i}}^{10}}+\ket{11}\ket{k_{A_{i}}^{11}}\nonumber\\
\ket{K_{B_{i}}}&=&\ket{00}\ket{k_{B_{i}}^{00}}+\ket{01}\ket{k_{B_{i}}^{01}}+\ket{10}\ket{k_{B_{i}}^{10}}+\ket{11}\ket{k_{B_{i}}^{11}}\nonumber\\
\ket{K_{AB_{i}}}&=&\ket{00}\ket{k_{AB_{i}}^{00}}+\ket{01}\ket{k_{AB_{i}}^{01}}+\ket{10}\ket{k_{AB_{i}}^{10}}+\ket{11}\ket{k_{AB_{i}}^{11}}.\nonumber
\end{eqnarray}
A product vector in  $\ket{e,f,g}\in V[\rho]$ has the property that it and its partial complex conjugates
have to be orthogonal to the corresponding kernels, i.e.:
\begin{eqnarray}\label{gsgen}
\braket{K_{i}}{e_{A},f_{B},g_{C}}&=&0,\nonumber\\
\braket{K_{A_{i}}}{e^{*}_{A},f_{B},g_{C}}&=&0,\nonumber\\
\braket{K_{B_{i}}}{e_{A},f^{*}_{B},g_{C}}&=&0,\nonumber\\
\braket{K_{AB_{i}}}{e^{*}_{A},f^{*}_{B},g_{C}}&=&0.
\end{eqnarray}
We expand now $\ket{e_{A},f_{B},g_{C}}$  in the local basis of  Alice and Bob:
\begin{eqnarray}
\ket{e_{A},f_{B},g_{C}}&=&(\alpha\ket{0}+\ket{1})\otimes (\beta\ket{0}+\ket{1})\otimes\ket{g}\nonumber\\
&=&(\alpha\beta\ket{00}+\alpha\ket{01}+\beta\ket{10}+\ket{11})\otimes\ket{g}.\nonumber
\end{eqnarray}
We observe that Eqs.  (\ref{gsgen}) can be rewritten as :
\begin{equation}
A(\alpha ,\beta ;\alpha^{*} ,\beta^{*})\ket{g}=0,
\end{equation} 
where $A(\alpha ,\beta;\alpha^{*} ,\beta^{*})$  is a
$(k(\rho)+k(\rho^{t_{A}})+k(\rho^{t_{B}})+k(\rho^{t_{AB}}))\times N$ matrix, which reads:
\begin{displaymath}
{\tiny{
A(\alpha ,\beta ;\alpha^{*} ,\beta^{*})=\left( \begin{array}{c}
\alpha\beta\bra{k_{i}^{00}}+\alpha\bra{k_{i}^{01}}+\beta\bra{k_{i}^{10}}+\bra{k_{i}^{11}}\\
\alpha^{*}\beta\bra{k_{A_{i}}^{00}}+\alpha^{*}\bra{k_{A_{i}}^{01}}+\beta\bra{k_{A_{i}}^{10}}+\bra{k_{A_{i}}^{11}}\\
\alpha\beta^{*}\bra{k_{B_{i}}^{00}}+\alpha\bra{k_{B_{i}}^{01}}+\beta^{*}\bra{k_{B_{i}}^{10}}+\bra{k_{B_{i}}^{11}}\\
\alpha^{*}\beta^{*}\bra{k_{AB_{i}}^{00}}+\alpha^{*}\bra{k_{AB_{i}}^{01}}+\beta^{*}\bra{k_{AB_{i}}^{10}}+\bra{k_{AB_{i}}^{11}}
\end{array}\right).}}
\end{displaymath}
Eqs. (\ref{gsgen}) have a nontrivial solution with $\ket{e}\not=0$,$\ket{f}\not=0$ and
 $\ket{g}\not=0$ iff  the rank of $A$ is smaller than  $N$. That implies that
at most $N-1$ rows  of the matrix $A$ are linearly independent. That means that
$(k(\rho)+k(\rho^{t_{A}})+k(\rho^{t_{B}})+k(\rho^{t_{AB}}))-N+1$ minors of dimension $N\times N$ of the
matrix 
$A$ must vanish. 

Let us consider the marginal case, when 
$k(\rho)+k(\rho^{t_{A}})+k(\rho^{t_{B}})+k(\rho^{t_{AB}})=2+(N-1)$.
In this case we combine the first $N-1$ rows with the remaining two and obtain exactly two different minors,
and thus two equations for complex $\alpha, \beta$, or more precisely four real equation for real and
imaginary parts of $\alpha,\beta$.  Such equations generically will have a finite number of solutions. 
The case when $k(\rho)+k(\rho^{t_{A}})+k(\rho^{t_{B}})+k(\rho^{t_{AB}})> 2+(N-1)$,  i.e. 
\begin{equation}
(r(\rho)+r(\rho^{t_{A}})+r(\rho^{t_{B}})+r(\rho^{t_{AB}})< 15N-1
\label{ineq}
\end{equation}
means that we have more equations than  parameters, and generically there will be no solution, or at least
the number of solutions will be even more limited than in the marginal case. 
The PPT states fulfilling the inequality (\ref{ineq}) are
generically the edge states, provided their rank and/or the ranks of their partial transposes are  greater that $N$, since
otherwise the Theorem of the previous section would apply.  Conversely, if 
\begin{equation}
(r(\rho)+r(\rho^{t_{A}})+r(\rho^{t_{B}})+r(\rho^{t_{AB}})> 15N-1, 
\label{ineq1}
\end{equation}
then the matrix $A$ has less equal than $N$ rows, and one can always use the freedom  of parameters to
find a solution, and subtract a projector onto a product vector from $\rho$ keeping its positivity 
and PPT property intact. 

In the following we will concentrate ourselves on the case  
$k(\rho)+k(\rho^{t_{A}})+k(\rho^{t_{B}})+k(\rho^{t_{AB}})\ge N+1$, for which the number of  solutions is
expected to be finite. Such states will be called as in Ref. \cite{MxN} generic. For those states it is simple
to check the separability, similarly as discussed in Ref. \cite{2xN,MxN}. The check is easy, because we know
that if the considered state is separable, then it is represented as a convex sum of projectors on the vectors
from the set $V[\rho]$, and the latter has a {\it finite} cardinality.  We will discuss this in more detail
for the case of  
${\cal C}^2\otimes{\cal C}^2\otimes{\cal C}^2$ systems.

\section{Separability checks  and criteria for generic low rank PPT states in  ${\cal C}^2\otimes{\cal
C}^2\otimes{\cal C}^2$ systems}

As a special, but important example we consider the case of 
PPT states in ${\cal C}^2\otimes{\cal C}^2\otimes{\cal
C}^2$ states  (3 qubit systems). We will use here the results of the previous sections. The 3 qubit case 
is particularly interesting as a first step toward multiple entangled systems, providing a challenge for 
both the theory and experiment. 

Generically, if 
\begin{equation}
(r(\rho)+r(\rho^{t_{A}})+r(\rho^{t_{B}})+r(\rho^{t_{AB}})\le 28, 
\label{ineq2}
\end{equation}
then the set $V[\rho]$ is empty and the state $\rho$ is a PPT entangled edge state, provided all the ranks
are greater 3, since otherwise the Lemma 5 of the previous section applies. We discuss the different cases 
below 
\subsection{The case $r(\rho)=2,3$}
From the results of the previous section we know that such PPT states are separable.

\subsection{The case $r(\rho)=4$}
The state of rank  $4$ in a 3 qubit system may be regarded a state in ${\cal C}^2\otimes{\cal
C}^4$ of rank 4. From the Theorem \ref{prodmn}  that this state is bipartite separable, and moreover has a
unique  decomposition into a sum of four projectors on product (biseparable) vectors in  $\quad{\cal
C}^{2}_{A}\otimes{\cal C}^{4}_{BC}$. From uniqueness, we gather that $\rho$ is then separable iff the
product vectors in this  decomposition are completely separable, i.e. are product vectors in 
${\cal C}^2\otimes{\cal C}^2\otimes{\cal
C}^2$.  Otherwise, the state is entangled, although biseparable. 
In fact it must be biseparable with respect to all partitions, i.e. also 
$\quad{\cal C}^{2}_{B}\otimes{\cal C}^{4}_{AC}$ and $\quad{\cal C}^{2}_{C}\otimes{\cal C}^{4}_{AB}$. 
Examples of such states are known, in particular those are the state constructed from unextendible product
basis \cite{upb}.

\subsection{The case $r(\rho)=r(\rho^{t_{A}})=5$}

This case is also easy because first of all the bipartite separability with respect to the partition $A-BC$
the  has to be checked. As shown in Ref. \cite{2xN}. a PPT state in $\quad{\cal
C}^{2}_{A}\otimes{\cal C}^{4}_{BC}$  is (bipartite) separable, iff the set of bipartite product vectors
$V_{A-BC}[\rho]$, corresponding to the partition $A-BC$ is not empty. In the present case it must not only
contain a bipartite product vector, but a tripartite product vector. If such vectors exist, generically
there will be finite number of them, and at least 5 of them must belong to the set $V_{A-B-C}[\rho]$.

\subsection{The case  $r(\rho)+r(\rho^{t_{A}})+r(\rho^{t_{B}})+ r(\rho^{t_{AB}})\le 28$}

In this case we have more equations than available parameters, and we expect that the set $V[\rho]$
will be empty, whereas the state $\rho$ will be an edge state. If this is not the case, we expect first of all
that there is  a finite number of product vectors in $V[\rho]$. Thus,  checking if $\rho$ can
be represented as a convex sum of projectors onto the elements of $V[\rho]$ can be performed exactly using the same 
methods as discussed in Ref. \cite{2xN}.

\subsection{The case $r(\rho)=r(\rho^{t_{A}})=r(\rho^{t_{B}})= r(\rho^{t_{AB}})=7$}

If $V[\rho]$ is empty, this case describes an example of an  
edge state with maximal sum of ranks. Such an
example has been constructed in Ref. \cite{ourclass}. Let us estimate  how many elements can 
the set $V[\rho]$ contain maximally. To this aim we write the matrix $A$:
\begin{displaymath}
{\tiny{
A(\alpha ,\beta ;\alpha^{*} ,\beta^{*})=\left( \begin{array}{c}
\alpha\beta\bra{k_{1}^{00}}+\alpha\bra{k_{1}^{01}}+\beta\bra{k_{1}^{10}}+\bra{k_{1}^{11}}\\
\alpha^{*}\beta\bra{k_{A_{1}}^{00}}+\alpha^{*}\bra{k_{A_{1}}^{01}}+\beta\bra{k_{A_{1}}^{10}}+\bra{k_{A_{1}}^{11}}\\
\alpha\beta^{*}\bra{k_{B_{1}}^{00}}+\alpha\bra{k_{B_{1}}^{01}}+\beta^{*}\bra{k_{B_{1}}^{10}}+\bra{k_{B_{1}}^{11}}\\
\alpha^{*}\beta^{*}\bra{k_{AB_{1}}^{00}}+\alpha^{*}\bra{k_{AB_{1}}^{01}}+\beta^{*}\bra{k_{AB_{1}}^{10}}+\bra{k_{AB_{1}}^{11}}
\end{array}\right)}}
\end{displaymath}
Let us denote  a polynomial $P$ of orders $X$ and $Y$ in variable  $z$ and $z^{*}$ by $P_{X,Y}(z)$. 
Combining the first and the third row, and the second and the fourth row   of $A$ we obtain the minors of the form:
{\footnotesize{
\begin{eqnarray}
\alpha^{2}P_{(1,1)}^{(1)}(\beta)+\alpha P_{(1,1)}^{(2)}(\beta)+P_{(1,1)}^{(3)}(\beta)=0.\label{m1}\\
(\alpha^{*})^2R_{(1,1)}^{(1)}(\beta)+\alpha^* R_{(1,1)}^{(2)}(\beta)+R_{(1,1)}^{(3)}(\beta)=0.\label{m3}
\end{eqnarray}}}
The remaining four combinations of rows give us 
{\footnotesize{
\begin{eqnarray}
\alpha\alpha^{*}Q_{(2,0)}^{(1)}(\beta)+\alpha Q_{(2,0)}^{(2)}(\beta)+\alpha^{*} Q_{(2,0)}^{(3)}
(\beta)+Q_{(2,0)}^{(4)}(\beta)=0\label{m2}\\
\alpha\alpha^{*}Q_{(0,2)}^{(1)}(\beta)+\alpha Q_{(0,2)}^{(2)}(\beta)+\alpha^{*} Q_{(0,2)}^{(3)}
(\beta)+Q_{(0,2)}^{(4)}(\beta)=0\label{m4},
\end{eqnarray}}}
and two equations of the form
{\footnotesize{
\begin{equation}
\alpha\alpha^{*}Q_{(1,1)}^{(1)}(\beta)+\alpha Q_{(1,1)}^{(2)}(\beta)+\alpha^{*} Q_{(1,1)}^{(3)}
(\beta)+Q_{(0,1)}^{(4)}(\beta)=0\label{m5}.
\end{equation}}}
Only 3 of the above equations are independent, but we have to our 
three complex conjugated equations to our disposal, and in
particular the conjugate of Eq. (\ref{m3}), 
{\footnotesize{
\begin{equation}
\alpha^2R_{(1,1)}^{(1)*}(\beta)+\alpha R_{(1,1)}^{(2)*}(\beta)+R_{(1,1)}^{(3)*}(\beta)=0.\label{m6}
\end{equation}}}
A good strategy is to  multiply Eq. (\ref{m1}) by $R_{(1,1)}^{(1)*}(\beta)$, and Eq. (\ref{m6}) by $P_{(1,1)}^{(1)}(\beta)$, and
subtract one from another in order to obtain
\begin{equation}
\alpha = T_{(2,2)}^{(1)}(\beta)/T_{(2,2)}^{(1)}(\beta).
\end{equation}
Inserting  this solution into Eq.  (\ref{m1}) we obtain a polynomial
 of orders  $5,5$ in  $\beta$ and
$\beta^{*}$.  Another independent polynomial
 is obtained by complex conjugation. The variables
$\beta$ and $\beta^*$ are then treated as independent ones, similarly in the Appendices of Ref.
\cite{2xN}. According to the result  presented there,
 a system of two polynomial equations of order $X,Y$ with $X\le Y$
for two variables $\beta$ and, say, $\bar\beta$ has at most 
$2^XY$ solutions for 
$\beta$. In the present case we expect thus that the number of solutions is  
$\le 160$. Most of these solutions  will have to be rejected typically, 
since they do not fulfill the conditions
 Eqs. (\ref{m2})-(\ref{m5}).

\subsection{The case  $r(\rho)+r(\rho^{t_{A}})+r(\rho^{t_{B}})+ r(\rho^{t_{AB}})= 29$}

This is a marginal case in which the number of equations is equal to the number of parameters, so that
generically we have a finite number of product vectors in $V[\rho]$, and a possibility of performing the
relatively straightforward separability check. For example,  if we consider
$r(\rho)=r(\rho^{t_{A}})= r(\rho^{t_{AB}})=7$ and $r(\rho^{t_{B}})=8$. In this case only two minors are independent, and we 
 have, for instance, 
to solve   Eq. (\ref{m2}), one of the Eqs. (\ref{m5}), 
and their complex conjugates. 
By multiplying Eqs. (\ref{m2})  
and (\ref{m5}) and its complex conjugates them by appropriate polynomials
in
$\beta,\beta^*$, and subtracting one from another we obtain 
two linear equations for $\alpha$, $\alpha^*$ of the form 
\begin{equation}\label{m7}
\alpha S_{(3,1)}^{(1)}(\beta)+\alpha^{*}S_{(3,1)}^{(2)}(\beta)+S_{(3,1)}^{(3)}(\beta)=0.
\end{equation}
and the complex conjugate of the above Eq. (\ref{m7}). 
This system of two linear equations can be
solved so that we obtain
\begin{equation}
\alpha = T_{(4,4)}^{(1)}(\beta)/T_{(4,4)}^{(1)}(\beta).
\end{equation}
Inserting this solution into Eq.  (\ref{m5}) and obtain in this way a 
polynomial of order  $9$ in  $\beta$ and
$\beta^{*}$. Another independent polynomial is obtained 
by complex conjugating Eq. (\ref{m2}). The variables
$\beta$ and $\beta^*$ are then treated as independent ones, 
similarly as discussed in the Appendix of Ref.
\cite{2xN}.   According to Ref. \cite{2xN} we expect in 
this case maximally $2^9\times 9=4608$ solutions
for $\beta$.

\subsection{Canonical form of non-decomposable entanglement witnesses}

For completeness it is worth mentioning that it is possible to 
generalize the results of Ref. \cite{witness}
to case of 3 qubit systems (and in general in tripartite systems). 
Let us remind the readers that
an entanglement witness is a hermitian operator $W$, for which 
${\rm Tr}(W\sigma)\ge 0$ for any
separable state $\sigma$, whereas ${\rm Tr}(W\rho)<0$ for some 
entangled state $\rho$. We say that $W$
detects then $\rho$. A non-decomposable witness is a witness that 
detects a PPT entangled state. Using exactly
the same arguments as in  Ref. \cite{witness}  one shows that a 
non-decomposable entanglement witness must have the
canonical form
\begin{eqnarray}
W=P+Q^{t_{A}}+R^{t_{B}}+ S^{t_{AB}}-\epsilon\eins,
\label{witness}
\end{eqnarray}
where 
\begin{equation}
\epsilon=\inf_{|e,f,g\rangle}\langle e,f,g|P+Q^{t_{A}}+R^{t_{B}}+ S^{t_{AB}}|e,f,g\rangle,
\end{equation}
the operators $P,Q,R,S$ are positively definite, $R(P)=K(\delta)$,  $R(Q)=K(\delta^{t_A})$,
$R(R)=K(\delta^{t_B})$, $R(S)=K(\delta^{t_{AB}})$, and  $\delta$ is an edge state, i.e. 
such state for which by
definition the set
$V[\delta]$ is empty, which implies automatically that $\epsilon$ is 
strictly positive. According to the 
results of this section, 
in three qubit system, the state  $\delta$ is a generic state with
$r(\delta)+r(\delta^{t_{A}})+r(\delta^{t_{B}})+ r(\delta^{t_{AB}})\le 28$.

\section{Conclusions}

We have generalized  previously obtained results for  PPT state in ${\cal C}^{2}\otimes{\cal C}^{N}$
and ${\cal C}^{M}\otimes{\cal C}^{N}$ system to PPT states in ${\cal C}^{2}\otimes{\cal C}^{2}\otimes{\cal
C}^{N}$. We have developed a method of "local projections" together with the PPT property to prove
separability of low rank states and to obtain separability criteria for low rank states.  These methods
together with methods developed in Refs. \cite{2xN,MxN} provide very general mathematical tools to study
separability and entanglement in multipartite systems.  

The main results of this paper are:
\begin{itemize}

\item The proof that all states with positive partial transposes 
that have rank $\le
N$ are separable, and have a certain canonical form;
\item The proof that  for the 3 qubit case ($N=2$) 
all PPT states
$\rho$ that have  rank $3$ are
separable; 
\item The presentation of constructive separability checks for the states  
$\rho$ that have the sum of the rank of $\rho$ and the ranks of partial
transposes with respect to all subsystems smaller than $15N-1$. 
\item The detailed discussion of the above mentioned 
constructive separability checks for the case $N=2$;
\item Presentation of the canonical form of
non-decomposable entanglement witnesses in 3 qubit systems. 

\end{itemize}

This work has been supported  by the DFG 
(SFB 407 and Schwerpunkt ``Quanteninformationsverarbeitung"), 
the ESF PESC Programm on Quantum Information,  and European Union IST Program
EQUIP.

\end{document}